\documentclass[aps,prl,superscriptaddress,twocolumn]{revtex4-1}
\usepackage{graphicx}
\usepackage{amsmath}
\usepackage{amssymb}
\usepackage{color}

\newcommand{\pd}{\partial}
\newcommand{\gr}{{Gr\"uneisen ratio }}

\begin{document}
\title{Finite-Size Scaling Theory at a Self-Dual Quantum Critical Point}

\author{Long Zhang}
\email{longzhang@ucas.ac.cn}
\affiliation{Kavli Institute for Theoretical Sciences and CAS Center for Excellence in Topological Quantum Computation, University of Chinese Academy of Sciences, Beijing 100190, China}

\author{Chengxiang Ding}
\affiliation{School of Microelectronics \& Data Science, Anhui University of Technology, Maanshan, Anhui 243002, China}


\date{\today}

\begin{abstract}

The nondivergence of the generalized \gr (GR) at a quantum critical point (QCP) has been proposed to be a universal thermodynamic signature of self-duality. In this work, we study how the Kramers-Wannier-type self-duality manifests itself in the finite-size scaling behavior of thermodynamic quantities in the quantum critical regime. While the self-duality cannot be realized as a unitary transformation in the total Hilbert space for the Hamiltonian with the periodic boundary condition, it can be implemented in certain symmetry sectors with proper boundary conditions. Therefore, the GR and the transverse magnetization of the one-dimensional transverse-field Ising model exhibit different finite-size scaling behaviors in different sectors. This implies that the numerical diagnosis of self-dual QCP requires identifying the proper symmetry sectors.

\end{abstract}
\maketitle

\emph{Introduction.---}Duality is a deep and elegant concept in statistical mechanics and quantum field theory \cite{Savit1980, Druhl1982}. It refers to the reformulation of a many-body system in terms of drastically different degrees of freedom, which typically maps a strongly coupled system into a weakly coupled one, thereby it facilitates further nonperturbative study. The recent discovery of duality between (2+1)-dimensional massless Dirac fermions and quantum electrodynamics (QED$_{3}$) offers a deep insight into the composite Fermi liquid in the half-filled lowest Landau level \cite{Son2015}.

Under a \emph{self-duality} transformation, the reformulated model has the same form as the original one but with a set of modified coupling constants. The Kramers-Wannier duality in the two-dimensional (2D) Ising model \cite{Kramers1941, Kadanoff1971} and its incarnation in the 1D transverse-field Ising model (TFIM) \cite{Fradkin1978} is the simplest example of self-duality. The spin order parameter is mapped into the nonlocal disorder parameter, and the paramagnetic phase into the ferromagnetic phase, and vice versa. The model at the critical point is invariant under the transformation. Therefore, the self-duality must pose stringent constraints on the universal critical phenomena.

In a recent work by one of the authors \cite{Zhang2019}, the non-divergence of the generalized \gr (GR) was proposed to be a universal thermodynamic signature of self-dual quantum critical points (QCPs). The GR is usually defined as the ratio of the thermal expansion coefficient $\alpha$ and the molar specific heat $c_{p}$, in which $\alpha=(1/V)(\pd V/\pd T)_{p}$, and $c_{p}=(\pd \epsilon/\pd T)_{p}$ with the molar energy density denoted by $\epsilon$. The GR can be generalized by replacing $p$ and $V$ with a pair of conjugate variables, i.e., a generalized force (a tuning parameter of the Hamiltonian) $g$ and the generalized displacement $v_{g}=(\pd f/\pd g)_{T}$,
\begin{equation}
\Gamma(T,g)=\frac{\pd_{T}v_{g}(T,g)}{\pd_{T}\epsilon(T,g)}. \label{eq:ggr}
\end{equation}
At a self-dual QCP with the tuning parameter $g$, the generalized GR was shown to be nondivergent in the zero-temperature limit, which is in sharp contrast to its universal power-law divergence at a QCP without self-duality \cite{Zhu2003}. This conclusion was drawn based on the hyperscaling theory in the quantum critical regime. The key point is the observation that the self-duality acts as a unitary or an antiunitary transformation in the Hilbert space and enforces the parity of the energy spectra on both sides of the QCP. This establishes the connection from the self-duality of the critical theory to the universal scaling function and observable quantities in critical phenomena. The nondivergent GR was first found in the 1DTFIM with its exact solution \cite{Wu2018} and observed in the quantum critical regime of a quasi-1D antiferromagnet BaCo$_{2}$V$_{2}$O$_{8}$ \cite{Wang2018a}.

In this work, we study the finite-size scaling theory of thermodynamic quantities at the self-dual QCPs, which is crucial for numerical calculations. The 1DTFIM is taken as an example to illustrate the main results. We first argue that the self-duality of the Kramers-Wannier-type, which maps local order parameters into nonlocal disorder operators, cannot be consistently defined as a unitary transformation with periodic boundary condition (PBC) in the total Hilbert space. This could be traced back to the nontrivial real-space topology \cite{Druhl1982}. This distinguishes the Kramers-Wannier-type self-duality from an internal $\mathbb{Z}_{2}$ symmetry, which is defined in terms of local degrees of freedom thus is not sensitive to the boundary conditions. Consequently, the GR of the 1DTFIM with the PBC at the QCP in the total Hilbert space is found to be divergent as the temperature $T\rightarrow 0$, if the product of $T$ and the lattice size $L$ is fixed.

How does the self-duality manifest itself \emph{on finite-size lattices}? It turns out that with proper boundary conditions, the self-duality can be defined as a unitary transformation in particular symmetry sectors, and the GR evaluated in such a sector does not diverge at the QCP. In the 1DTFIM, the self-duality is well-defined in the $\mathbb{Z}_{2}$-even sector for the PBC, and the $\mathbb{Z}_{2}$-odd sector for the anti-periodic boundary condition (APBC). More generically, the self-duality can be defined in the total Hilbert space if the Hamiltonian is supplemented with extra boundary terms. In other words, the Hamiltonian can be decomposed into an exactly self-dual part, plus an extra boundary term. The extra boundary contribution to the thermodynamic quantities on finite-size lattices is found to be suppressed as $y=1/(T^{1/z}L)\rightarrow 0$, in which $z$ is the dynamical exponent, thus the nondivergence of the GR at the self-dual QCP is recovered in the thermodynamic limit. Besides, the self-duality also leaves traces in the finite-size corrections to the transverse magnetization at the QCP, which can also be used as numerical evidence of self-duality.

\begin{figure*}[t]
\includegraphics[width=\textwidth]{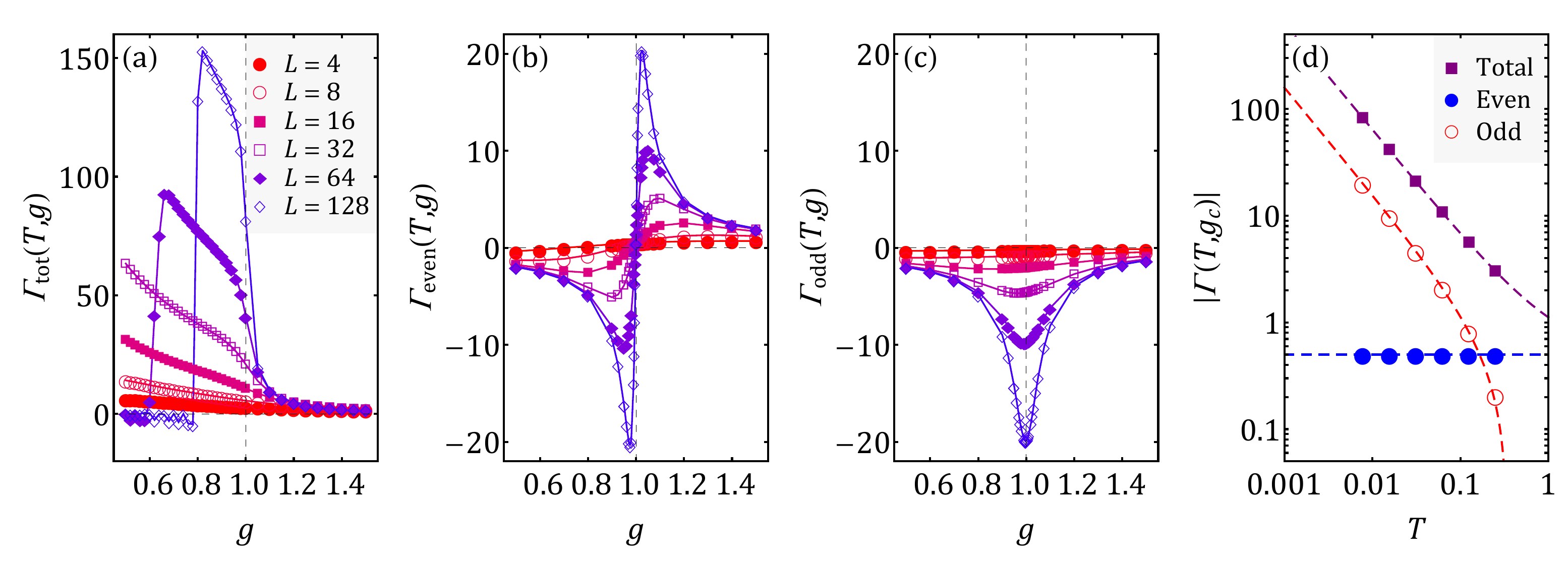}
\caption{The \gr (GR) of the one-dimensional transverse-field Ising model (1DTFIM) close to the quantum critical point (QCP) on finite-size lattices with the periodic boundary condition (PBC) evaluated in (a) the total Hilbert space, (b) the $\mathbb{Z}_{2}$-even sector, and (c) the $\mathbb{Z}_{2}$-odd sector, respectively. The temperature $T=1/L$. (d) The GR at the QCP versus $T$ in the double-log scale. The GR diverges as $T^{-1}$ in the total Hilbert space and in the $\mathbb{Z}_{2}$-odd sector, but remains precisely $1/2$ independent of $T$ in the $\mathbb{Z}_{2}$-even sector.}
\label{fig:tfi}
\end{figure*}

\emph{Self-duality of 1DTFIM on finite-size lattices.---}The 1DTFIM is defined by
\begin{equation}
H(g)=-\sum_{i}\sigma_{i}^{z}\sigma_{i+1}^{z}-g\sum_{i}\sigma_{i}^{x},
\end{equation}
in which $\sigma_{i}^{z}$ and $\sigma_{i}^{x}$ are Pauli matrices. This model has an internal $\mathbb{Z}_{2}$ symmetry under the action of $\Sigma=\prod_{i}\sigma_{i}^{x}$. The Kramers-Wannier self-duality is usually formulated in terms of the operator mapping \cite{Fradkin1978},
\begin{equation}
\sigma_{i}^{z}\mapsto \prod_{l\leq i}\sigma_{l}^{x},\quad \sigma_{i}^{x}\mapsto \sigma_{i}^{z}\sigma_{i+1}^{z}, \label{eq:mapping}
\end{equation}
which preserves the anticommutation relation of the Pauli matrices. The Hamiltonian $H(g)$ is mapped into $gH(1/g)$, thus the QCP at $g_{c}=1$ is self-dual. The exact critical exponents are $z=1$ and $\nu=1$.

For the 1DTFIM, the generalized displacement $v_{g}=-m_{x}(T,g)$, in which $m_{x}(T,g)=(1/L)\sum_{i}\langle\sigma_{i}^{x}\rangle_{T}$ is the transverse magnetization at temperature $T$. The generalized GR $\Gamma(T,g)$ defined in Eq. (\ref{eq:ggr}) is reduced to
\begin{equation}
\Gamma(T,g)=-\frac{\partial_{T}m_{x}(T,g)}{\partial_{T}\epsilon(T,g)}. \label{eq:gr}
\end{equation}
The GR at the QCP $\Gamma(T,g_{c})$ was shown to be nondivergent as $T\rightarrow 0$ if the self-duality acts as a unitary or an antiunitary transformation \cite{Zhang2019}. This behavior distinguishes a self-dual QCP from generic cases, for which the GR diverges with a power-law, $\Gamma(T,g_{c})\propto T^{-1/z\nu}$, in which $\nu$ is the correlation-length exponent \cite{Zhu2003}.

However, in the 1DTFIM on a finite-size lattice with PBC, the self-duality mapping of operators cannot be consistently realized as a unitary transformation in the total Hilbert space. With the operator mapping in Eq. (\ref{eq:mapping}), the $\mathbb{Z}_{2}$ symmetry operator $\Sigma$ would be mapped to $\prod_{i}\sigma_{i}^{z}\sigma_{i+1}^{z}$, which equals the identity operator in the PBC because $(\sigma_{i}^{z})^{2}=1$. This implies that the operator mapping is not self-consistent. Similar issues arise in the generalization of the Kramers-Wannier duality to other lattice models with Abelian symmetry \cite{Savit1980, Druhl1982}.

Let us first show that the self-duality can be defined as a unitary transformation in certain symmetry sectors with proper boundary conditions, based on which the finite-size scaling of the free energy is derived in the next section.

The total Hilbert space is the direct sum of two eigen-subspaces of the symmetry operator $\Sigma$ with eigenvalues $+1$ and $-1$, which are called the $\mathbb{Z}_{2}$-even and the $\mathbb{Z}_{2}$-odd sectors, respectively. The $\mathbb{Z}_{2}$-even sector is spanned by
\begin{equation}
|s_{1},s_{2},\ldots,s_{N}\rangle_{+}=\frac{1}{\sqrt{2}}(1+\Sigma)\otimes_{i}|s_{i}\rangle_{z}, \label{eq:basis}
\end{equation}
which is obtained by applying the projection operator $P_{+}=(1+\Sigma)/2$ to the direct-product state in the $\sigma_{i}^{z}$ basis, $\sigma_{i}^{z}|s_{i}\rangle_{z}=s_{i}|s_{i}\rangle_{z}$, and an extra $\sqrt{2}$ factor is to maintain the normalization. The self-duality transformation $U_{+}$ in this sector is given by
\begin{equation}
U_{+}|s_{1},s_{2},\ldots,s_{N}\rangle_{+} =\otimes_{i}|s_{i-1}s_{i}\rangle_{x}, \label{eq:upbc}
\end{equation}
in which $|s_{i}\rangle_{x}$ denotes the $\sigma_{i}^{x}$ basis with $\sigma_{i}^{x}|s_{i}\rangle_{x}=s_{i}|s_{i}\rangle_{x}$ and is related to the $\sigma_{i}^{z}$ basis by $|s_{i}\rangle_{x}=(|1\rangle_{z}+s_{i}{|-1\rangle_{z}})/\sqrt{2}$. It is easy to check that the state on the right-hand side of Eq. (\ref{eq:upbc}) is in the $\mathbb{Z}_{2}$-even sector as well, thus this sector is closed under the $U_{+}$ action. The $U_{+}$ transformation of operators restricted in this sector are obtained by examining their action on the basis given in Eq. (\ref{eq:basis}). For example,
\begin{equation}
U_{+}\sigma_{i}^{x}|s_{1},s_{2},\ldots,s_{N}\rangle_{+} = \sigma_{i}^{z}\sigma_{i+1}^{z}U_{+}|s_{1},s_{2},\ldots,s_{N}\rangle_{+},
\end{equation}
thus $U_{+}\sigma_{i}^{x}U_{+}^{-1} = \sigma_{i}^{z}\sigma_{i+1}^{z}$. Similarly, $U_{+}\sigma_{i}^{z}\sigma_{i+1}^{z}U_{+}^{-1}=\sigma_{i+1}^{x}$. Therefore, the Hamiltonian with the PBC restricted in this sector $H_{+}(g)$ is exactly self-dual, $UH_{+}(g)U^{-1}=gH_{+}(1/g)$. Moreover, applying $U_{+}$ twice results in a spatial translation, $U_{+}^{2}=T_{x}$, in which $T_{x}$ acts as $T_{x}|s_{1},s_{2},\ldots,s_{N}\rangle_{+}=|s_{N},s_{1},\ldots,s_{N-1}\rangle_{+}$.

The $\mathbb{Z}_{2}$-odd sector is spanned by
\begin{equation}
|s_{1},s_{2},\ldots,s_{N}\rangle_{-} =\frac{s_{1}}{\sqrt{2}}(1-\Sigma)|s_{1},s_{2},\ldots,s_{N}\rangle_{z},
\end{equation}
which is obtained by applying the projection operator $P_{-}=(1-\Sigma)/2$ to the direct-product state in the $\sigma_{z}$ basis, and the extra $s_{1}$ is to guarantee the consistency, $|s_{1},s_{2},\ldots,s_{N}\rangle_{-}=|-s_{1},-s_{2},\ldots,-s_{N}\rangle_{-}$. In this sector, the unitary self-duality transformation can be defined by
\begin{equation}
U_{-}|s_{1},s_{2},\ldots,s_{N}\rangle_{-}=|-s_{N}s_{1}\rangle_{x}\otimes|s_{1}s_{2}\rangle_{x}\otimes\cdots\otimes|s_{N-1}s_{N}\rangle_{x}. \label{eq:uapbc}
\end{equation}
The operators in this sector transform as $U_{-}\sigma_{i}^{x}U_{-}^{-1}=\eta_{i}\sigma_{i}^{z}\sigma_{i+1}^{z}$, and $U_{-}\sigma_{i}^{z}\sigma_{i+1}^{z}U_{-}^{-1}=\eta_{i}\sigma_{i+1}^{x}$, in which the sign factor $\eta_{i}$ equals $-1$ for $i=N$ and is $+1$ otherwise. The Hamiltonian with the APBC restricted in this sector $H_{-}(g)$ is also exactly self-dual, $U_{-}H_{-}(g)U_{-}^{-1}=gH_{-}(1/g)$. Similarly, we find $U_{-}^{2}=T_{x}$, in which $T_{x}|s_{1},s_{2},\ldots,s_{N}\rangle_{-}=|s_{N},s_{1},\ldots,s_{N-1}\rangle_{-}$.

Therefore, the proper definition of the self-duality as a unitary transformation depends on the boundary condition and the symmetry sector. This is a generic feature of Kramers-Wannier-type dualities of lattice gauge theory and spin models with Abelian symmetry, because the partition function on a lattice with nontrivial topology is mapped into a weighted sum over the partition functions with different boundary conditions \cite{Druhl1982}.

The total Hilbert space is the direct sum of the $\mathbb{Z}_{2}$-even and the $\mathbb{Z}_{2}$-odd sectors, thus the unitary transformation $U=U_{+}\oplus U_{-}$ is the exact self-duality transformation for $H_{\mathrm{SD}}=H_{+}\oplus H_{-}$. The 1DTFIM with PBC is given by $H_{\mathrm{SD}}$ plus an extra boundary operator, $H=H_{\mathrm{SD}}+2P_{-}\sigma_{N}^{z}\sigma_{1}^{z}P_{-}$, in which $P_{-}$ is the projection operator into the $\mathbb{Z}_{2}$-odd sector.

The 1DTFIM can be exactly solved with the Jordan-Wigner transformation \cite{Pfeuty1970, Zhang2015a}, $\sigma_{i}^{z}=\prod_{l<i}(2c_{l}^{\dagger}c_{l}-1)(c_{i}+c_{i}^{\dagger})$ and $\sigma_{i}^{x}=2c_{i}^{\dagger}c_{i}-1$, in which $c_{i}$ and $c_{i}^{\dagger}$ are fermion operators. The Hamiltonian is mapped to free fermions,
\begin{equation}
H(g)=-\sum_{i}(c_{i}-c_{i}^{\dagger})(c_{i+1}^{\dagger}+c_{i+1})-g\sum_{i}(2c_{i}^{\dagger}c_{i}-1).
\end{equation}
The eigenstates of the 1DTFIM can be constructed by filling the single-particle eigenstates. The $\mathbb{Z}_{2}$ symmetry corresponds to the fermion parity, $\Sigma=\prod_{i}(2c_{i}^{\dagger}c_{i}-1)$. For large system sizes, the full summation over the eigenstates is prohibitively costly, thus we adopt the Metropolis Monte Carlo sampling over the single-particle occupation numbers to evaluate the thermodynamic quantities.

The GR of the 1DTFIM with PBC are evaluated in the total Hilbert space and in each symmetry sector, i.e., the thermodynamic average is taken within each sector \cite{Pfeuty1970, Zhang2015a}. The results are shown in Fig. \ref{fig:tfi}. In the $\mathbb{Z}_{2}$-even sector, the GR at the QCP $\Gamma_{\mathrm{even}}(T,g_{c})$ is precisely $1/2$ at any $T$ due to the exact self-duality in this sector \cite{Zhang2019}. In contrast, the GR evaluated in the total Hilbert space and in the $\mathbb{Z}_{2}$-odd sector diverge as $T^{-1}$ ($z\nu=1$ in 1DTFIM) on finite lattices with $L= T^{-1}$. This is surprising at first sight, given that the exact self-duality is only broken by an extra boundary term, whose impact is expected to vanish in the thermodynamic limit. This paradox is resolved in the finite-size scaling theory presented in the following section.

\begin{figure*}[tb]
\includegraphics[width=\textwidth]{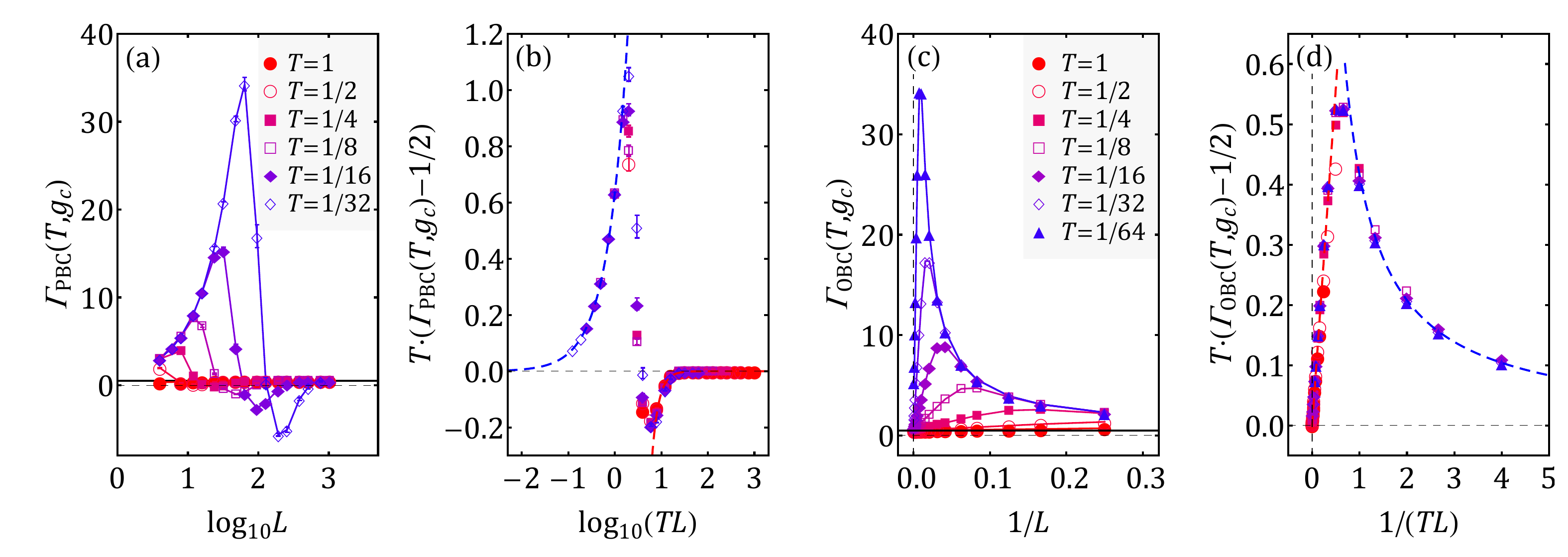}
\caption{The GR of the 1DTFIM at the QCP on finite-size lattices with (a) the PBC and (c) the open boundary condition (OBC). By rescaling the axes, we find perfect data collapse in (b) and (d) in a form given by $\Gamma(T,g_{c},L)-1/2\simeq T^{-1}\tilde{\Gamma}(y)$, in which $y=1/(TL)$. A $1/2$ is subtracted from $\Gamma(T,g_{c},L)$ to take care of the subleading nondivergent part. The red dashed lines are the power-law fitting in the $y\rightarrow 0$ limit: $\tilde{\Gamma}(y)\propto y^{3}$ in (b) and $\tilde{\Gamma}(y)\propto y$ in (d). The blue dashed lines are the fitting of $\tilde{\Gamma}(y)\propto y^{-1}$ in the $y\rightarrow \infty$ limit, which comes from the low-energy excitations in the $T\ll 1/L$ limit.}
\label{fig:tdl}
\end{figure*}

\emph{Finite-size scaling of thermodynamic quantities.---}Close to the QCP, the singular part of the free energy density has the finite-size scaling form \cite{Barber1983phase}
\begin{equation}
f_{s}(T,g,L)\simeq T^{d/z+1}\tilde{f}_{s}\bigg(\frac{\delta g}{T^{1/z\nu}},\frac{1}{T^{1/z}L}\bigg), \label{eq:fss-f}
\end{equation}
in which $d$ is the spatial dimension, and $\delta g=g-g_{c}$. The reduced variables are denoted by $x=\delta g/T^{1/z\nu}$ and $y=1/(T^{1/z}L)$ in the rest of this work. $\tilde{f}_{s}(x,y)$ is a universal scaling function, which reduces to the scaling function $\Phi(x)$ introduced in Refs. \cite{Zhu2003, Zhang2019} in the thermodynamic limit.

The singular part of the energy density $\epsilon_{s}(T,g,L)$ and the transverse magnetization $m_{x,s}(T,g,L)$ close to the QCP can be derived from Eq. (\ref{eq:fss-f}),
\begin{gather}
\epsilon_{s}(T,g,L) \simeq -T^{d/z+1}\Big(\frac{d}{z}-\frac{1}{z\nu}x\partial_{x}-\frac{1}{z}y\partial_{y}\Big)\tilde{f}_{s}(x,y),\\
m_{x,s}(T,g,L) \simeq -T^{d/z+1-1/z\nu}\partial_{x}\tilde{f}_{s}(x,y).
\end{gather}
Therefore, the finite-size scaling of the GR in the quantum critical regime, where $|x|\ll 1$, is given by
\begin{equation}
\Gamma(T,g,L) =-\frac{\partial_{T} m_{x,s}}{\partial_{T} \epsilon_{s}} \simeq -T^{-1/z\nu}\tilde{\Gamma}(y), \label{eq:fss-gamma}
\end{equation}
in which
\begin{equation}
\tilde{\Gamma}(y) = \frac{z\big((d+z-1/\nu)\partial_{x}\tilde{f}_{s}(0,y)-y\partial_{x,y}^{2}\tilde{f}_{s}(0,y)\big)}{(d+z-y\partial_{y})(d-y\partial_{y})\tilde{f}_{s}(0,y)}.
\end{equation}
It reduces to the scaling form $\Gamma(T,g)\propto -G_{T}T^{-1/z\nu}$ obtained in Ref. \cite{Zhu2003} as $y\rightarrow 0$ in the thermodynamic limit.

What is the impact of self-duality on the finite-size scaling of $\Gamma(T,g_{c},L)$? Let us show that the nondivergence of $\Gamma(T,g_{c})$ as $T\rightarrow 0$ at a self-dual QCP is recovered if the thermodynamic limit is taken properly. As shown in the previous section, the Hamiltonian $H(g)$ near a self-dual QCP can be written as an exactly self-dual part $H_{\mathrm{SD}}(g)$ plus an extra boundary term $B(g)$ up to irrelevant operators, $H(g)=H_{\mathrm{SD}}(g)+B(g)$, thus the free energy $f(T,g,L)$ of $H(g)$ differs from that of $H_{\mathrm{SD}}(g)$ by a boundary term,
\begin{equation}
f(T,g,L)=f_{\mathrm{SD}}(T,g,L)+f_{B}(T,g,L).
\end{equation}
In the finite-size scaling of $f_{\mathrm{SD}}(T,g,L)$ in the quantum critical regime, $f_{\mathrm{SD}}(T,g,L)\simeq T^{d/z+1}\tilde{f}_{\mathrm{SD}}(x,y)$, and the scaling function $\tilde{f}_{\mathrm{SD}}(x,y)$ satisfies $\partial_{x}\tilde{f}_{\mathrm{SD}}(0,y)=0$ for any $y$, because the self-duality acts as a unitary transformation on any finite-size lattices for the Hamiltonian $H_{\mathrm{SD}}(g)$. The excess contribution from the boundary satisfies \cite{Binder1983Phase}
\begin{equation}
f_{B}(T,g,L)\simeq \frac{\xi}{L}(\xi_{\tau}\xi^{d})^{-1}\tilde{f}_{B}(x,y)\propto T^{d/z+1}y\tilde{f}_{B}(x,y),
\end{equation}
in which $\xi\propto T^{-1/z}$ and $\xi_{\tau}\propto T^{-1}$ are the spatial and the temporal correlation lengths in the quantum critical regime, respectively. Therefore, $f_{B}$ is suppressed by a factor of $y$ as compared with $f_{\mathrm{SD}}$. The scaling function $\tilde{f}_{s}(x,y)=\tilde{f}_{\mathrm{SD}}(x,y)+y\tilde{f}_{B}(x,y)$, which implies that in the quantum critical regime,
\begin{equation}
\tilde{\Gamma}(y)=\frac{z(d+z-1/\nu)\partial_{x}\tilde{f}_{\mathrm{SD}}(0,y)}{d(d+z)\tilde{f}_{\mathrm{SD}}(0,y)}+O(y)\rightarrow 0, \label{eq:y0}
\end{equation}
as $y\rightarrow 0$. The boundary contribution vanishes if the thermodynamic limit is taken by sending $y\rightarrow 0$.

The GR of the 1DTFIM at the QCP calculated with the periodic and the open boundary conditions are shown in Fig. \ref{fig:tdl}, which exhibit perfect data collapse consistent with the scaling form in Eq. (\ref{eq:fss-gamma}). The scaling function $\tilde{\Gamma}(y)$ is not identically zero, thus $\Gamma(T,g_{c},L)\propto T^{-1/z\nu}$ as $T\rightarrow 0$ with $y$ fixed at a finite value. However, we find $\tilde{\Gamma}(y)\rightarrow 0$ as $y\rightarrow 0$, which is compatible with Eq. (\ref{eq:y0}). The numerical results of the PBC suggests an even faster decay, $\tilde{\Gamma}(y)\propto y^{3}$, because the boundary term $B=2P_{-}\sigma_{N}^{z}\sigma_{1}^{z}P_{-}$ is nonzero only in the $\mathbb{Z}_{2}$-odd sector, thus its contribution to the free energy is further suppressed by the excitation gap.

\begin{figure}[tb]
\includegraphics[width=\columnwidth]{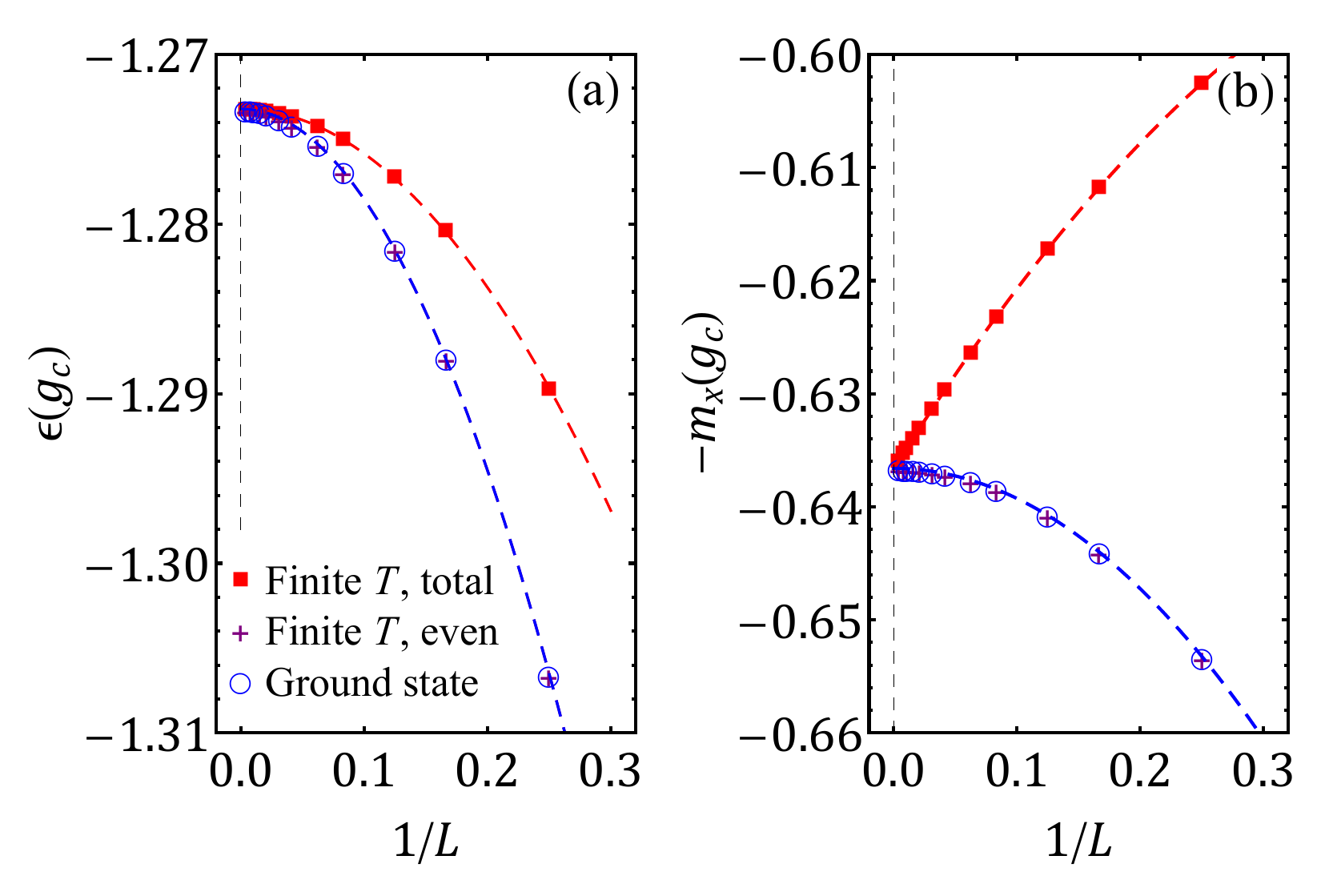}
\caption{(a) The energy density and (b) the transverse magnetization of the 1DTFIM on finite-size lattices with the PBC, which are evaluated in the total Hilbert space (red squares) and in the $\mathbb{Z}_{2}$-even sector (purple crosses) at $T=1/L$, and at the ground state (blue circles). The finite-size correction to the energy density scales as $L^{-2}$ in all three cases, while the correction to the transverse magnetization scales as $L^{-1}$ in the total Hilbert space and as $L^{-2}$ in the other two cases, for which the self-duality is well-defined.}
\label{fig:energy}
\end{figure}

\emph{Finite-size corrections to energy and transverse magnetization.---}The information of the self-duality can also be inferred from the finite-size corrections to the transverse magnetization at the QCP, because the leading-order correction comes from the singular part of the free energy and is not changed by irrelevant perturbations. We consider the scheme of taking the thermodynamic limit with a fixed $y=1/(T^{1/z}L)$, which is usually adopted in numerical calculations of quantum critical phenomena.

The finite-size correction to the energy density at the QCP scales as $\delta\epsilon(T,g_{c},L)\propto T^{d/z+1}\propto L^{-(d/z+1)}$, which is valid no matter whether the QCP is self-dual. The results for the 1DTFIM are shown in Fig. \ref{fig:energy} (a).

On the other hand, the finite-size correction form of the transverse magnetization is modified by the self-duality. For a generic QCP, the leading-order correction scales as
\begin{equation}
\begin{split}
\delta m_{x}(T,g_{c},L) &\simeq  -T^{d/z+1-1/z\nu}\partial_{x}\tilde{f}_{s}(0,y) \\
& \propto L^{-(d/z+1-1/z\nu)}\partial_{x}\tilde{f}_{s}(0,y).
\end{split}
\label{eq:mx}
\end{equation}
The prefactor $\partial_{x}\tilde{f}_{s}(0,y)$ is nonzero for a non-self-dual QCP as well as for the 1DTFIM with the PBC averaged in the total Hilbert space [see Fig. \ref{fig:energy} (b)]. In contrast, the prefactor vanishes for the 1DTFIM with the PBC if the average is taken in the $\mathbb{Z}_{2}$-even sector or on the ground state, where the self-duality is a well-defined unitary transformation, thus the leading-order correction to $m_{x}(T,g,L)$ comes in higher order, $\delta m_{x}(T,g_{c},L)\propto L^{-(d/z+1)}$ due to the smooth variation of the free energy with $g$.

\emph{Summary and discussions.---}In summary, we have shown how the Kramers-Wannier-type self-duality manifests itself in the finite-size scaling of thermodynamic quantities. In order to define the self-duality as a unitary transformation, the boundary condition and the symmetry sector must be properly specified. In the 1DTFIM, the self-duality cannot be consistently defined in the total Hilbert space for the Hamiltonian with the PBC; Instead, it can be defined in the $\mathbb{Z}_{2}$-even sector with the PBC or in the $\mathbb{Z}_{2}$-odd sector with the APBC. Consequently, the GR at the QCP evaluated in the total Hilbert space diverges as $T\rightarrow 0$ for fixed $y=1/(T^{1/z}L)$, $\Gamma(T,g_{c},L)\simeq \tilde{\Gamma}(y)T^{-1/z\nu}$; Nevertheless, the prefactor $\tilde{\Gamma}(y)$ vanishes as $y\rightarrow 0$. The GR evaluated in the $\mathbb{Z}_{2}$-even sector with the PBC does not diverge as $T\rightarrow 0$ due to the well-defined self-duality transformation. Besides, the finite-size corrections to the transverse magnetization evaluated in the total Hilbert space and in the $\mathbb{Z}_{2}$-even sector exhibit different power-law behavior.

For a generic QCP with Kramers-Wannier-type self-duality, the self-duality is well-defined as a unitary transformation only in the charge-neutral sector with PBC, i.e., the Hilbert subspace carrying the trivial representation of the Abelian symmetry. The generalized GR $\Gamma(T,g,L)$ and the generalized displacement $v_{g}(T,g,L)$ follow the same finite-size scaling form as Eqs. (\ref{eq:fss-gamma}) and (\ref{eq:mx}). Moreover, the scaling behavior is not changed by irrelevant perturbation. Therefore, we could numerically diagnose the possible Kramers-Wannier-type self-duality at the QCP in the proper symmetry sector.

\begin{acknowledgments}
L.Z. thanks Shenghan Jiang, Ya-Wen Sun and Huajia Wang for helpful discussions. L.Z. is supported by National Key R\&D Program of China (No. 2018YFA0305800), National Natural Science Foundation of China (Nos. 12174387 and 11804337), Strategic Priority Research Program of CAS (No. XDB28000000), and the CAS Youth Innovation Promotion Association. C.D. is supported by National Natural Science Foundation of China (Nos. 11975024 and 62175001), and the Anhui Provincial Supporting Program for Excellent Young Talents in Colleges and Universities under Grant No. gxyqZD2019023.
\end{acknowledgments}

\bibliography{library,books}

\begin{thebibliography}{14}%
\makeatletter
\providecommand \@ifxundefined [1]{%
 \@ifx{#1\undefined}
}%
\providecommand \@ifnum [1]{%
 \ifnum #1\expandafter \@firstoftwo
 \else \expandafter \@secondoftwo
 \fi
}%
\providecommand \@ifx [1]{%
 \ifx #1\expandafter \@firstoftwo
 \else \expandafter \@secondoftwo
 \fi
}%
\providecommand \natexlab [1]{#1}%
\providecommand \enquote  [1]{``#1''}%
\providecommand \bibnamefont  [1]{#1}%
\providecommand \bibfnamefont [1]{#1}%
\providecommand \citenamefont [1]{#1}%
\providecommand \href@noop [0]{\@secondoftwo}%
\providecommand \href [0]{\begingroup \@sanitize@url \@href}%
\providecommand \@href[1]{\@@startlink{#1}\@@href}%
\providecommand \@@href[1]{\endgroup#1\@@endlink}%
\providecommand \@sanitize@url [0]{\catcode `\\12\catcode `\$12\catcode
  `\&12\catcode `\#12\catcode `\^12\catcode `\_12\catcode `\%12\relax}%
\providecommand \@@startlink[1]{}%
\providecommand \@@endlink[0]{}%
\providecommand \url  [0]{\begingroup\@sanitize@url \@url }%
\providecommand \@url [1]{\endgroup\@href {#1}{\urlprefix }}%
\providecommand \urlprefix  [0]{URL }%
\providecommand \Eprint [0]{\href }%
\providecommand \doibase [0]{http://dx.doi.org/}%
\providecommand \selectlanguage [0]{\@gobble}%
\providecommand \bibinfo  [0]{\@secondoftwo}%
\providecommand \bibfield  [0]{\@secondoftwo}%
\providecommand \translation [1]{[#1]}%
\providecommand \BibitemOpen [0]{}%
\providecommand \bibitemStop [0]{}%
\providecommand \bibitemNoStop [0]{.\EOS\space}%
\providecommand \EOS [0]{\spacefactor3000\relax}%
\providecommand \BibitemShut  [1]{\csname bibitem#1\endcsname}%
\let\auto@bib@innerbib\@empty
\bibitem [{\citenamefont {Savit}(1980)}]{Savit1980}%
  \BibitemOpen
  \bibfield  {author} {\bibinfo {author} {\bibfnamefont {R.}~\bibnamefont
  {Savit}},\ }\href@noop {} {\bibfield  {journal} {\bibinfo  {journal} {Rev.
  Mod. Phys.}\ }\textbf {\bibinfo {volume} {52}},\ \bibinfo {pages} {453}
  (\bibinfo {year} {1980})}\BibitemShut {NoStop}%
\bibitem [{\citenamefont {Dr{\"{u}}hl}\ and\ \citenamefont
  {Wagner}(1982)}]{Druhl1982}%
  \BibitemOpen
  \bibfield  {author} {\bibinfo {author} {\bibfnamefont {K.}~\bibnamefont
  {Dr{\"{u}}hl}}\ and\ \bibinfo {author} {\bibfnamefont {H.}~\bibnamefont
  {Wagner}},\ }\href {\doibase 10.1016/0003-4916(82)90286-X} {\bibfield
  {journal} {\bibinfo  {journal} {Ann. Phys. (N. Y).}\ }\textbf {\bibinfo
  {volume} {141}},\ \bibinfo {pages} {225} (\bibinfo {year}
  {1982})}\BibitemShut {NoStop}%
\bibitem [{\citenamefont {Son}(2015)}]{Son2015}%
  \BibitemOpen
  \bibfield  {author} {\bibinfo {author} {\bibfnamefont {D.~T.}\ \bibnamefont
  {Son}},\ }\href {\doibase 10.1103/PhysRevX.5.031027} {\bibfield  {journal}
  {\bibinfo  {journal} {Phys. Rev. X}\ }\textbf {\bibinfo {volume} {5}},\
  \bibinfo {pages} {031027} (\bibinfo {year} {2015})}\BibitemShut {NoStop}%
\bibitem [{\citenamefont {Kramers}\ and\ \citenamefont
  {Wannier}(1941)}]{Kramers1941}%
  \BibitemOpen
  \bibfield  {author} {\bibinfo {author} {\bibfnamefont {H.~A.}\ \bibnamefont
  {Kramers}}\ and\ \bibinfo {author} {\bibfnamefont {G.~H.}\ \bibnamefont
  {Wannier}},\ }\href {\doibase 10.1103/PhysRev.60.252} {\bibfield  {journal}
  {\bibinfo  {journal} {Phys. Rev.}\ }\textbf {\bibinfo {volume} {60}},\
  \bibinfo {pages} {252} (\bibinfo {year} {1941})}\BibitemShut {NoStop}%
\bibitem [{\citenamefont {Kadanoff}\ and\ \citenamefont
  {Ceva}(1971)}]{Kadanoff1971}%
  \BibitemOpen
  \bibfield  {author} {\bibinfo {author} {\bibfnamefont {L.~P.}\ \bibnamefont
  {Kadanoff}}\ and\ \bibinfo {author} {\bibfnamefont {H.}~\bibnamefont
  {Ceva}},\ }\href {\doibase 10.1103/PhysRevB.3.3918} {\bibfield  {journal}
  {\bibinfo  {journal} {Phys. Rev. B}\ }\textbf {\bibinfo {volume} {3}},\
  \bibinfo {pages} {3918} (\bibinfo {year} {1971})}\BibitemShut {NoStop}%
\bibitem [{\citenamefont {Fradkin}\ and\ \citenamefont
  {Susskind}(1978)}]{Fradkin1978}%
  \BibitemOpen
  \bibfield  {author} {\bibinfo {author} {\bibfnamefont {E.}~\bibnamefont
  {Fradkin}}\ and\ \bibinfo {author} {\bibfnamefont {L.}~\bibnamefont
  {Susskind}},\ }\href@noop {} {\bibfield  {journal} {\bibinfo  {journal}
  {Phys. Rev. D}\ }\textbf {\bibinfo {volume} {17}},\ \bibinfo {pages} {2637}
  (\bibinfo {year} {1978})}\BibitemShut {NoStop}%
\bibitem [{\citenamefont {Zhang}(2019)}]{Zhang2019}%
  \BibitemOpen
  \bibfield  {author} {\bibinfo {author} {\bibfnamefont {L.}~\bibnamefont
  {Zhang}},\ }\href {\doibase 10.1103/PhysRevLett.123.230601} {\bibfield
  {journal} {\bibinfo  {journal} {Phys. Rev. Lett.}\ }\textbf {\bibinfo
  {volume} {123}},\ \bibinfo {pages} {230601} (\bibinfo {year}
  {2019})}\BibitemShut {NoStop}%
\bibitem [{\citenamefont {Zhu}\ \emph {et~al.}(2003)\citenamefont {Zhu},
  \citenamefont {Garst}, \citenamefont {Rosch},\ and\ \citenamefont
  {Si}}]{Zhu2003}%
  \BibitemOpen
  \bibfield  {author} {\bibinfo {author} {\bibfnamefont {L.}~\bibnamefont
  {Zhu}}, \bibinfo {author} {\bibfnamefont {M.}~\bibnamefont {Garst}}, \bibinfo
  {author} {\bibfnamefont {A.}~\bibnamefont {Rosch}}, \ and\ \bibinfo {author}
  {\bibfnamefont {Q.}~\bibnamefont {Si}},\ }\href {\doibase
  10.1103/PhysRevLett.91.066404} {\bibfield  {journal} {\bibinfo  {journal}
  {Phys. Rev. Lett.}\ }\textbf {\bibinfo {volume} {91}},\ \bibinfo {pages}
  {066404} (\bibinfo {year} {2003})}\BibitemShut {NoStop}%
\bibitem [{\citenamefont {Wu}\ \emph {et~al.}(2018)\citenamefont {Wu},
  \citenamefont {Zhu},\ and\ \citenamefont {Si}}]{Wu2018}%
  \BibitemOpen
  \bibfield  {author} {\bibinfo {author} {\bibfnamefont {J.}~\bibnamefont
  {Wu}}, \bibinfo {author} {\bibfnamefont {L.}~\bibnamefont {Zhu}}, \ and\
  \bibinfo {author} {\bibfnamefont {Q.}~\bibnamefont {Si}},\ }\href {\doibase
  10.1103/PhysRevB.97.245127} {\bibfield  {journal} {\bibinfo  {journal} {Phys.
  Rev. B}\ }\textbf {\bibinfo {volume} {97}},\ \bibinfo {pages} {245127}
  (\bibinfo {year} {2018})}\BibitemShut {NoStop}%
\bibitem [{\citenamefont {Wang}\ \emph {et~al.}(2018)\citenamefont {Wang},
  \citenamefont {Lorenz}, \citenamefont {Gorbunov}, \citenamefont {Cong},
  \citenamefont {Kohama}, \citenamefont {Niesen}, \citenamefont {Breunig},
  \citenamefont {Engelmayer}, \citenamefont {Herman}, \citenamefont {Wu},
  \citenamefont {Kindo}, \citenamefont {Wosnitza}, \citenamefont {Zherlitsyn},\
  and\ \citenamefont {Loidl}}]{Wang2018a}%
  \BibitemOpen
  \bibfield  {author} {\bibinfo {author} {\bibfnamefont {Z.}~\bibnamefont
  {Wang}}, \bibinfo {author} {\bibfnamefont {T.}~\bibnamefont {Lorenz}},
  \bibinfo {author} {\bibfnamefont {D.~I.}\ \bibnamefont {Gorbunov}}, \bibinfo
  {author} {\bibfnamefont {P.~T.}\ \bibnamefont {Cong}}, \bibinfo {author}
  {\bibfnamefont {Y.}~\bibnamefont {Kohama}}, \bibinfo {author} {\bibfnamefont
  {S.}~\bibnamefont {Niesen}}, \bibinfo {author} {\bibfnamefont
  {O.}~\bibnamefont {Breunig}}, \bibinfo {author} {\bibfnamefont
  {J.}~\bibnamefont {Engelmayer}}, \bibinfo {author} {\bibfnamefont
  {A.}~\bibnamefont {Herman}}, \bibinfo {author} {\bibfnamefont
  {J.}~\bibnamefont {Wu}}, \bibinfo {author} {\bibfnamefont {K.}~\bibnamefont
  {Kindo}}, \bibinfo {author} {\bibfnamefont {J.}~\bibnamefont {Wosnitza}},
  \bibinfo {author} {\bibfnamefont {S.}~\bibnamefont {Zherlitsyn}}, \ and\
  \bibinfo {author} {\bibfnamefont {A.}~\bibnamefont {Loidl}},\ }\href
  {\doibase 10.1103/PhysRevLett.120.207205} {\bibfield  {journal} {\bibinfo
  {journal} {Phys. Rev. Lett.}\ }\textbf {\bibinfo {volume} {120}},\ \bibinfo
  {pages} {207205} (\bibinfo {year} {2018})}\BibitemShut {NoStop}%
\bibitem [{\citenamefont {Pfeuty}(1970)}]{Pfeuty1970}%
  \BibitemOpen
  \bibfield  {author} {\bibinfo {author} {\bibfnamefont {P.}~\bibnamefont
  {Pfeuty}},\ }\href@noop {} {\bibfield  {journal} {\bibinfo  {journal} {Ann.
  Phys. (N. Y).}\ }\textbf {\bibinfo {volume} {57}},\ \bibinfo {pages} {79}
  (\bibinfo {year} {1970})}\BibitemShut {NoStop}%
\bibitem [{\citenamefont {Zhang}\ and\ \citenamefont
  {Song}(2015)}]{Zhang2015a}%
  \BibitemOpen
  \bibfield  {author} {\bibinfo {author} {\bibfnamefont {G.}~\bibnamefont
  {Zhang}}\ and\ \bibinfo {author} {\bibfnamefont {Z.}~\bibnamefont {Song}},\
  }\href {\doibase 10.1103/PhysRevLett.115.177204} {\bibfield  {journal}
  {\bibinfo  {journal} {Phys. Rev. Lett.}\ }\textbf {\bibinfo {volume} {115}},\
  \bibinfo {pages} {177204} (\bibinfo {year} {2015})}\BibitemShut {NoStop}%
\bibitem [{\citenamefont {Barber}(1983)}]{Barber1983phase}%
  \BibitemOpen
  \bibfield  {author} {\bibinfo {author} {\bibfnamefont {M.~N.}\ \bibnamefont
  {Barber}},\ }in\ \href@noop {} {\emph {\bibinfo {booktitle} {Phase
  Transitions and Critical Phenomena}}},\ Vol.~\bibinfo {volume} {8},\ \bibinfo
  {editor} {edited by\ \bibinfo {editor} {\bibfnamefont {C.}~\bibnamefont
  {Domb}}\ and\ \bibinfo {editor} {\bibfnamefont {J.~L.}\ \bibnamefont
  {Lebowitz}}}\ (\bibinfo  {publisher} {Academic Press},\ \bibinfo {address}
  {London, England},\ \bibinfo {year} {1983})\BibitemShut {NoStop}%
\bibitem [{\citenamefont {Binder}(1983)}]{Binder1983Phase}%
  \BibitemOpen
  \bibfield  {author} {\bibinfo {author} {\bibfnamefont {K.}~\bibnamefont
  {Binder}},\ }in\ \href@noop {} {\emph {\bibinfo {booktitle} {Phase
  Transitions and Critical Phenomena}}},\ Vol.~\bibinfo {volume} {8},\ \bibinfo
  {editor} {edited by\ \bibinfo {editor} {\bibfnamefont {C.}~\bibnamefont
  {Domb}}\ and\ \bibinfo {editor} {\bibfnamefont {J.~L.}\ \bibnamefont
  {Lebowitz}}}\ (\bibinfo  {publisher} {Academic Press},\ \bibinfo {address}
  {London, England},\ \bibinfo {year} {1983})\BibitemShut {NoStop}%
\end{thebibliography}%
\end{document}